\documentclass[runningheads]{svmult}

\usepackage{makeidx}   
\usepackage{graphicx}  
\usepackage{subeqnar}  
\usepackage{multicol}  
\usepackage{physprbb}  
\makeindex             

\newcommand{\kms}{\,km~s$^{-1}$}      

\newcommand{\msun}{\,{\rm M_\odot}}
\newcommand{\ltsima}{$\; \buildrel < \over \sim \;$}
\newcommand{\simlt}{\lower.5ex\hbox{\ltsima}}
\newcommand{\gtsima}{$\; \buildrel > \over \sim \;$}
\newcommand{\simgt}{\lower.5ex\hbox{\gtsima}}

\begin{document}
\title*{Star Forming Galaxies in the `Redshift Desert'
}
\toctitle{Star-Forming Galaxies in the `Redshift Desert'}
%
%
\titlerunning{Galaxies in the Redshift Desert}
%
\author{C. Steidel\inst{1}
\and A. Shapley\inst{2}
\and M. Pettini\inst{3}
\and K. Adelberger\inst{4}
\and D. Erb\inst{1}
\and N. Reddy\inst{1}
\and M. Hunt\inst{1}
}
\authorrunning{Steidel et al.}
%
%
\institute{California Institute of Technology, MS 105-24, \\
Pasadena, CA 91125, USA 
\and University of California, Berkeley
\and Institute of Astronomy, Cambridge, UK
\and Carnegie Observatories}                      

\maketitle              

\begin{abstract}
We describe results of optical and near-IR observations of a large
spectroscopic sample of star-forming galaxies photometrically-selected to lie in the redshift
range $1.4 \simlt z \simlt 2.5$, often called the ``redshift desert''
because of historical difficulty in obtaining spectroscopic redshifts
in this range. We show that the former ``redshift desert'' is now very much
open to observation. 
\end{abstract}

\section{Background}

The ``redshift desert'' results from an accident of nature in which
the windows of low atmospheric opacity and low terrestrial background 
are barren of familiar, strong spectroscopic features that make redshift
identification easy using ground-based spectroscopy. At $z \sim 1.4$ the last 
of the strong nebular lines, [OII] $\lambda 3727$, passes well into the range
where most optical spectrographs perform less well because of decreasing
CCD quantum efficiency and rapidly increasing sky brightness. The result
has been that there is a dearth of direct spectroscopic information on
galaxies at $z \simgt 1.4$ until $z \simgt 2.5$, at which point (for star
forming galaxies, at least) techniques like Lyman break selection coupled with
normal optical spectroscopy have been quite successful. The difficulties with
spectroscopy in the desert translate directly into larger uncertainties in
photometric redshifts, since both methods depend on strong spectral features 
at observationally accessible wavelengths. 
Of course, nature has conspired
to make this redshift range between $z \sim 1.5-2.5$ perhaps one of the most 
crucial in understanding the development of massive galaxies and of their
central black holes, as we have heard from a number of talks at this meeting. 
There is thus a very strong impetus to gain access to galaxies in this range
of redshifts, in spite of the difficulties that may be encountered.

Following the familiar rest-optical nebular lines into the near-IR is certainly possible in principle, but
to date there has been little effort to do this in a wholesale manner because
there have not been many instruments capable of multiplexed spectroscopy in the near-IR;
this of course will change significantly over the next $\sim 3-5$ years as cryogenic
multi-object near-IR spectrographs come on line.  
We have adopted an alternative strategy: to push faint object spectroscopy into the blue/UV
portion of the observed spectrum, where the same spectral features used for the identification
of galaxies at $z \sim 3$ \cite{s03} remain accessible down to $z \sim 1.4$, thereby closing the gap
in redshift space that has been called the ``desert''. 

Studying galaxies at $z \sim 2$ is quite rewarding for both scientific and practical reasons:
This redshift range $1.5 \simlt z \simlt 2.5$ evidently contains the peak of the QSO epoch, 
and may well contain the formation era for most of the stars in today's massive galaxies, 
judging by the redshift distribution of bright sub-mm sources \cite{chap03}. But it is
much more than a crucial epoch for galaxies and AGN-- at these redshifts, it is possible
to simultaneously study the diffuse intergalactic medium (IGM) {\it and} star forming
galaxies in the same volumes, allowing for a direct examination of the magnitude of the effects of
supernova feedback on the properties of galaxies and the IGM. Because both background QSOs
and spectroscopically accessible galaxies have much higher surface densities than at higher
redshifts where such experiments have already been done \cite{A03}, $z \sim2$ may be the optimal
redshift for joint IGM/galaxy studies. In terms of attaining physical understanding of the
galaxies one finds, $z \sim 2$ offers significant advantages as well: first, the surface density
of galaxies bright enough for detailed spectroscopic studies using 8m-class telescopes
at both optical and near-IR wavelengths is high; secondly, as we describe below, one has
access to diagnostic spectroscopy in both the rest-frame far-UV {\it and} the
rest-frame optical, allowing independent means of measuring physical properties such as chemical abundances,
stellar initial mass function, and mass. 

\section{Survey of the ``Redshift Desert''}

We began our survey of $z \sim 2$ galaxies in the fall of 2000 just after the
commissioning of the LRIS-B instrument (see \cite{s04}) on the Keck I telescope.
Our approach to identifying which galaxies to target in order to efficiently
survey galaxies at $z \sim 2$ was largely empirical, in the sense that we 
chose the region in color-space that would be occupied by galaxies having
the same rest-frame SEDs as $z \sim 3$ LBGs in the sample of objects that have
been observed from $0.3-2.2\mu$m (see \cite{A04} for a full description and
motivation of the color selection). Generally speaking, this means that the
galaxies we are selecting for spectroscopy have the same range of UV color
as the better-studied LBGs at $z \sim 3$. Because we began our survey with the primary
aim of studying the connection between the galaxies and the IGM, we have focused
most of our efforts on the redshift range $2 \simlt z \simlt 2.5$ (we refer to
this particular color selection as ``BX'' objects) where the Lyman
$\alpha$ forest is easily studied from the ground, and in fields having more than
1 suitable $z\sim2.5$ background QSO for studying the IGM component. However, we
have recently obtained additional spectroscopy in both the GOODS-N and Groth/Westphal
fields, where in addition to the $z \sim 2-2.5$ ``BX'' objects we have targeted
objects expected to lie in the range $1.5 \simlt z \simlt 2$ (and which
we refer to as ``BM'' objects). A more in-depth overview of the survey and its initial results is
given in \cite{s04}.

\begin{figure}[tb]
\begin{center}
\includegraphics[width=.8\textwidth]{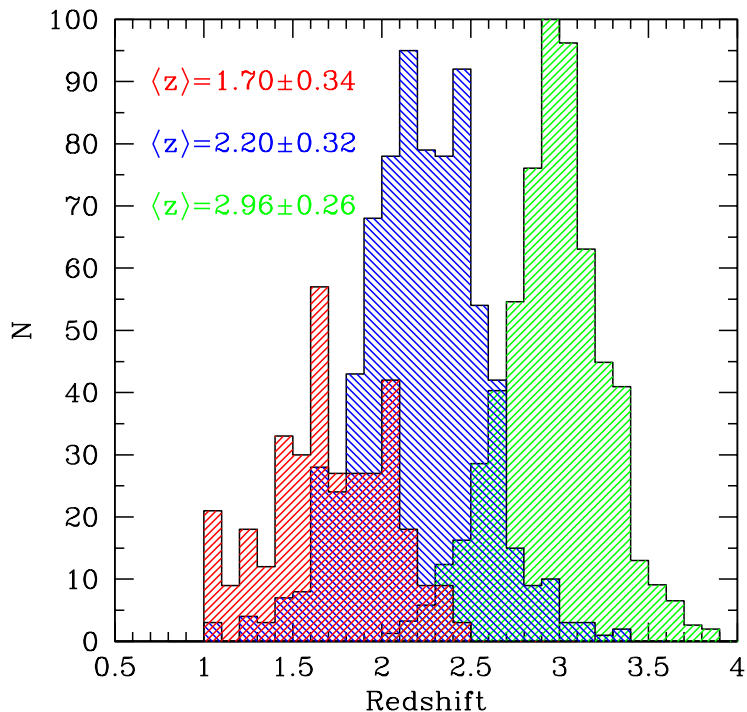}
\end{center}
\caption[]{Redshift histograms for rest-UV selected samples of star forming 
galaxies. The blue histogram is the 750 ``BX'' galaxies with spectroscopic
redshifts at the time of this writing; the red histogram results from the
``BM'' selection targeting $z\simeq 1.5-2.0$ (118 redshifts, scaled by 3 for
plotting). The green histogram is the $z \sim 3$ LBG sample of \cite{s03} (940 redshifts,
scaled by 0.7 for plotting). The total
number of new redshifts in the range $1.4 \le z \le 2.6$ is 792 at this time.}  
\label{zhist}
\end{figure}

Figure 1 shows the results to date for the spectroscopy of the two new photometrically
selected samples, with $\langle z \rangle = 2.20\pm 0.32$ and $\langle z \rangle = 1.70\pm0.34$
for the ``BX'' and ``BM'' samples, respectively. Figure 2 shows example LRIS-B spectra obtained
since the science-grade detector system was installed in the instrument in June 2002. 

The location of the $z=1.5-2.5$ galaxies in color space (we use the same 3-band system
used to isolate $z \sim 3$ galaxies) is easily estimated \cite{A04}, and in fact galaxies in
this range of redshifts are very common. After correcting for $\sim$ 8\%  contamination
by stars and low-redshift galaxies in the photometric samples, the total surface
density of BX+BM galaxies is $\sim 9$ arcmin$^{-2}$ to ${\cal R}=25.5$ (our adopted
spectroscopic limit), or about 25\% of the number counts to the same apparent magnitude
limit. With only 90-minute total integration times, the LRIS-B spectroscopy achieves
$\sim 70\%$ success in identifying redshifts, primarily based on numerous interstellar
absorption lines, as illustrated in Figure 2. Because LRIS-B is so efficient in the UV-visual
range, and the sky is extremely dark, it is possible to collect high quality continuum spectroscopy
in very reasonable integration times, so that the detection of strong emission lines is
completely unnecessary for achieving high spectroscopic completeness. At the time of this
writing, we have obtained a total of more than 790 spectroscopic redshifts in the range
$1.4 \simlt z \simlt 2.6$, with the distributions as shown in Figure 1. 
We have had similar success using LRIS-B to obtain spectra of objects which failed
to yield redshifts in the DEEP2 redshift survey (see Jeff Newman's contribution to
these proceedings), where because of the spectroscopic configuration used, it is difficult
to measure redshifts at $z > 1.4$. In general, LRIS-B is ideal for measuring redshifts for just the
objects that generally fail to yield redshifts for spectroscopic setups focused on
the red part of the optical spectrum.

\begin{figure}[tb]
\begin{center}
\includegraphics[width=0.5\textwidth]{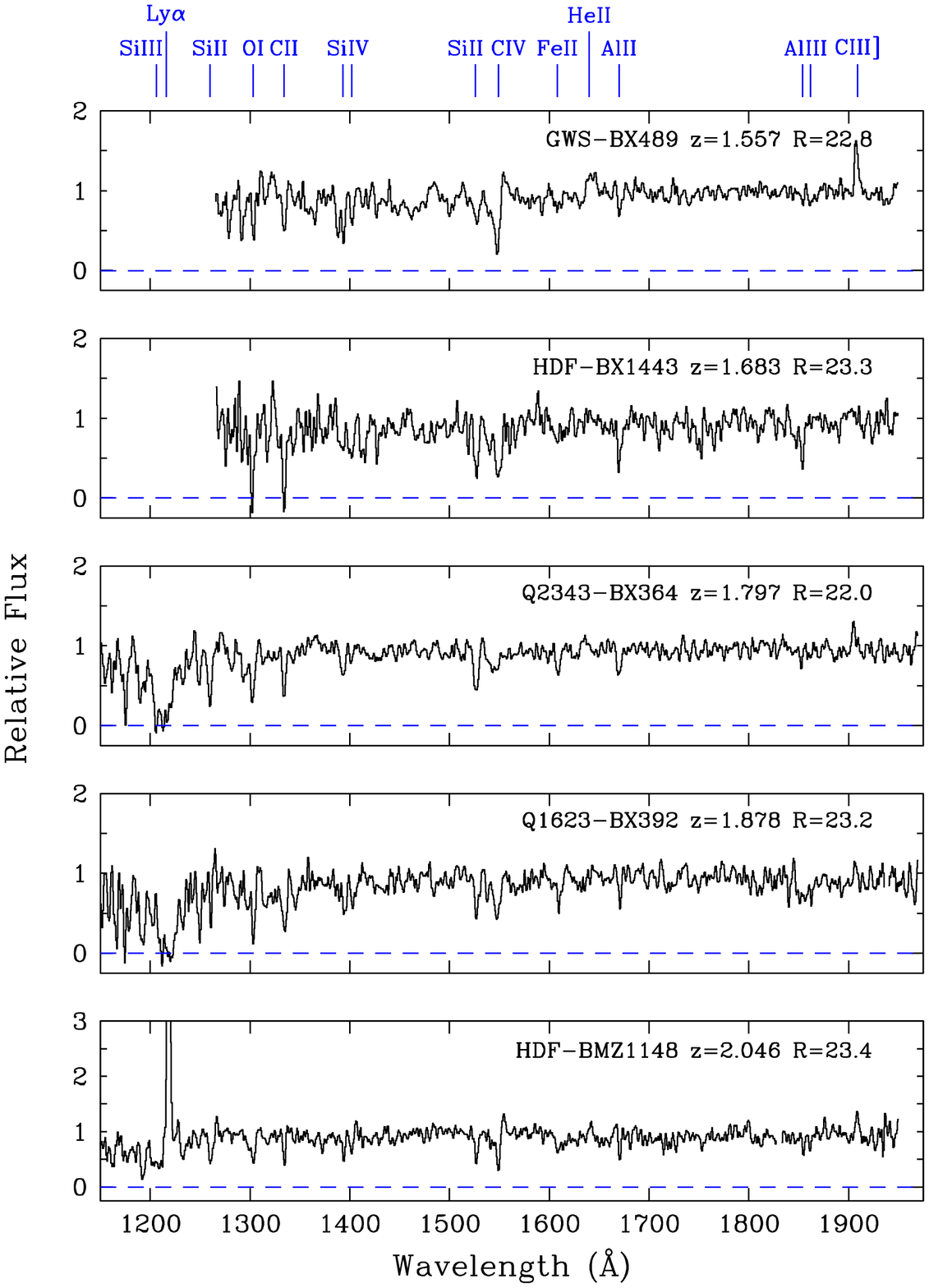}\includegraphics[width=0.5\textwidth]{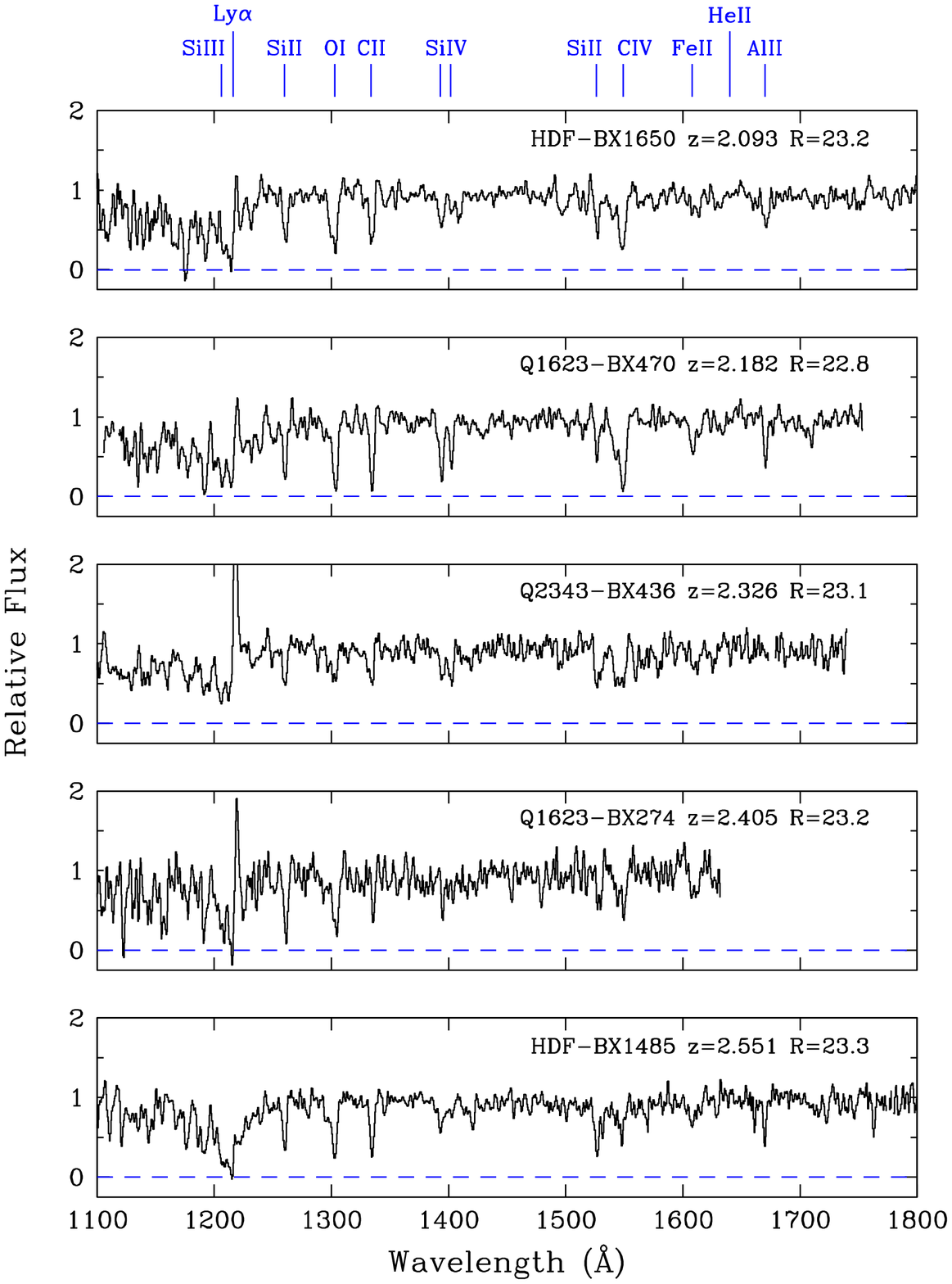}
\end{center}
\caption[]{Example  LRIS-B spectra (shifted into the rest frame, but generally covering the wavelength
range $3200-6000$\AA) of galaxies in the BX/BM sample covering the full range of redshifts and
spectral properties seen in the sample. All spectra were obtained with 90-minute total integration
times, in survey mode. The spectral resolution is $\sim 1.5-2.0$ \AA\ in the rest frame.} 
\label{spec}
\end{figure}

\section{Near-IR Observations of Optically Selected Desert Galaxies}

\subsection{Imaging}

It is interesting to investigate the rest-frame optical properties of the rest-UV-selected
``desert'' galaxy population, particularly as a means of comparing them differentially
to the similarly-selected $z \sim 3$ sample, and to various other surveys that use
near-IR selection (e.g., FIRES, K20, GDDS--see contributions from Labb\'e, Cimatti, 
and Chen in these proceedings, respectively ). Beginning in June 2003, with the
commissioning of the Wide Field Infrared Camera (WIRC) on the Palomar 5m telescope, we
have been obtaining very deep near-IR images of the regions of our survey fields with
the densest spectroscopic coverage. In the first 3 pointings of the camera, which
has a field of view of $8.7 \times 8.7$ arcmin, we have obtained $K_s$ band measurements
for a total of 283 `BX' galaxies that already have spectroscopic redshifts. Although it
requires very long ($\sim 12$ hour) integrations on a 5m telescope, we have been
able to reach limiting magnitudes of $K \sim 22.3$, allowing us to detect $\sim 85$\% of
the galaxies with spectra that fall within each pointing. Figure 3 shows a color-magnitude diagram
of the 283 detected galaxies in the current sample. About 8\% of these galaxies have $K_s < 20$
(the limit for the K20 survey) and $\sim 30$\% has $K_s < 20.6$ (the limit for the Gemini
Deep Deep Survey).

\begin{figure}[tb]
\begin{center}
\includegraphics[width=0.5\textwidth]{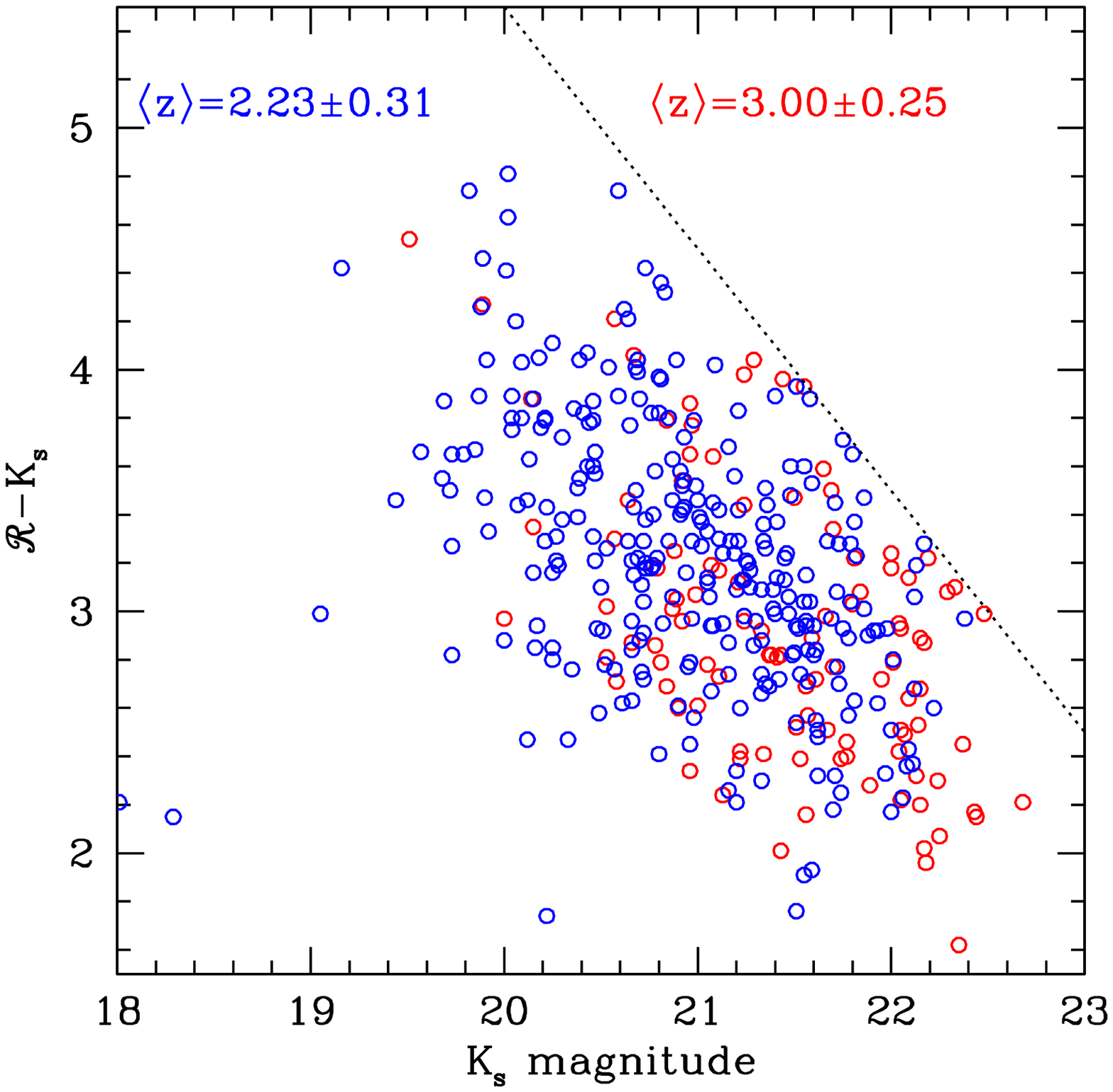}\includegraphics[width=0.5\textwidth]{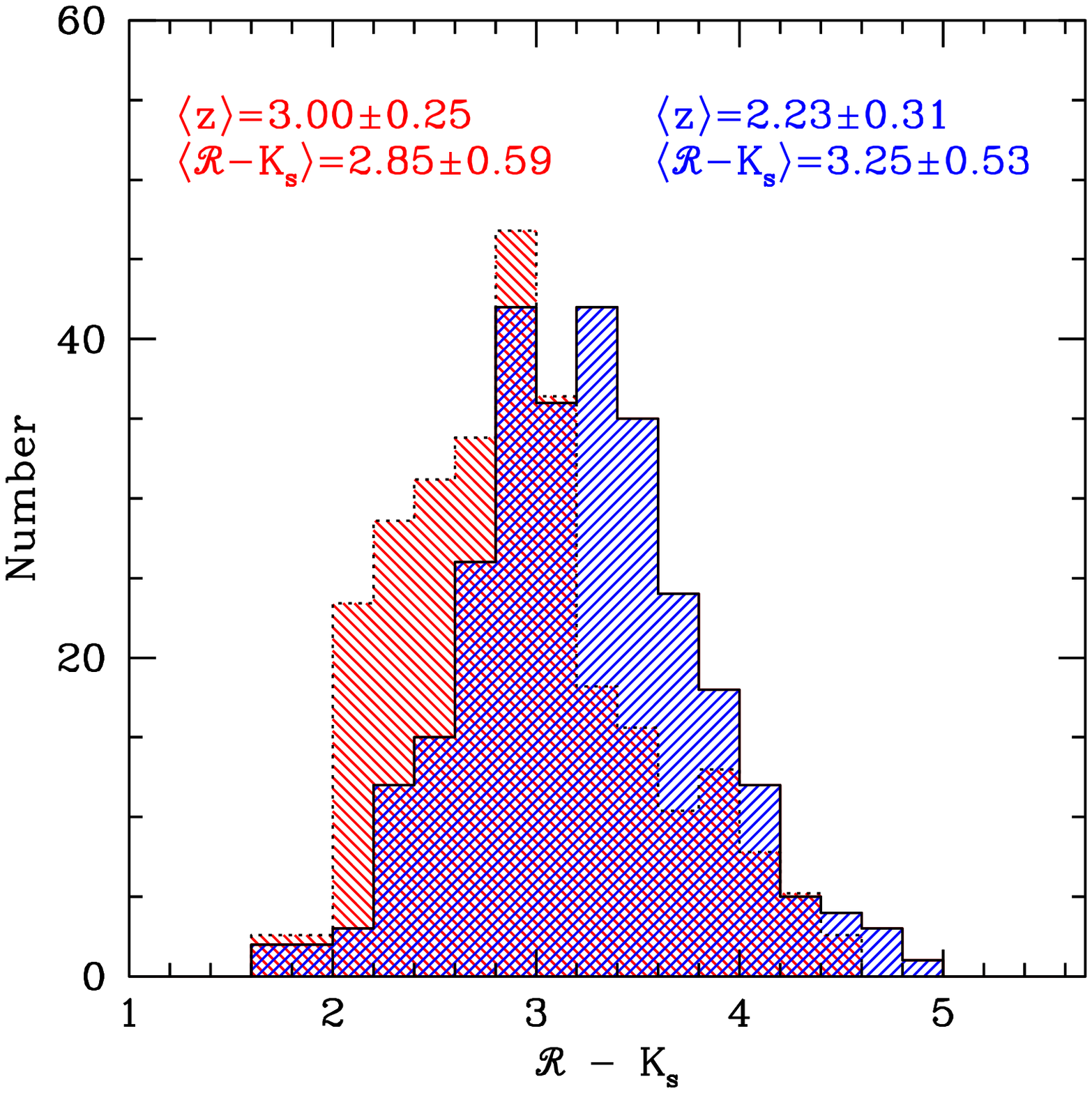}
\end{center}
\caption[]{{\bf Left:} Color/magnitude diagram showing the 283 $z \sim 2$ galaxies with K-band
measurements in our spectroscopic sample (blue points). The red points are taken from
the $z \sim3$ sample of \cite{shap01}. The dotted line corresponds
to our spectroscopic limit (for both surveys) of ${\cal R}=25.5$. {\bf Right:} Histograms
of optical/IR color for the two samples, normalized to the same numbers. The $z \sim 2$
sample has many more galaxies of with (${\cal R}-K_s \simgt 3.2$), which tend to be objects
with significant Balmer breaks indicating extended star formation histories.} 
\label{kband}
\end{figure}

A cursory comparison of the distribution of ${\cal R}-K_s$ colors of the samples
of star forming galaxies at $ z\sim 3$ and $z \sim 2$ indicates that the $z \sim 2$ galaxies
are significantly redder. Since objects with identical star formation histories observed
at $z \sim 3$ and $z \sim 2$ are predicted to have essentially identical ${\cal R}-K_s$
color, the result suggests a real difference in the rest-optical properties 
of objects selected using the same rest-UV criteria in the two redshift intervals. 
The most likely cause of this is an increase in the average stellar mass of star forming
galaxies at $z \sim 2$ compared to their counterparts at $z \sim 3$. These results will
be quantified in future work. 

\subsection{Spectroscopy}

Galaxies at $z \sim 2$, and especially those at redshifts $z=2-2.5$,
are extremely well-suited for follow-up spectroscopy in the near-IR because
the various nebular lines fall at favorable redshifts with respect to the
atmospheric J, H, and K band windows. Using primarily the NIRSPEC instrument
on the Keck II telescope,  we have focused on observing
the H$\alpha$ line primarily in the K-band window (initial results are
presented in \cite{erb03} and in the poster paper presented by Dawn Erb at this
meeting). The scientific aims of these observations are manifold, and include
1) accurate measurement of the galaxy systemic redshift, crucial for analysis
of the galaxy/IGM interface and for evaluating the velocities of supernova-driven
outflows in the galaxies 2) measurement of the galaxy kinematics (line widths,
rotation curves, etc.) 3) measuring chemical abundances in the galaxy's HII
regions using various nebular line diagnostics and 4) additional estimates of star formation rates
in the galaxies, possibly less affected by extinction than estimates based on UV continuum
measurements.

\begin{figure}[tb]
\begin{center}
\includegraphics[width=0.5\textwidth]{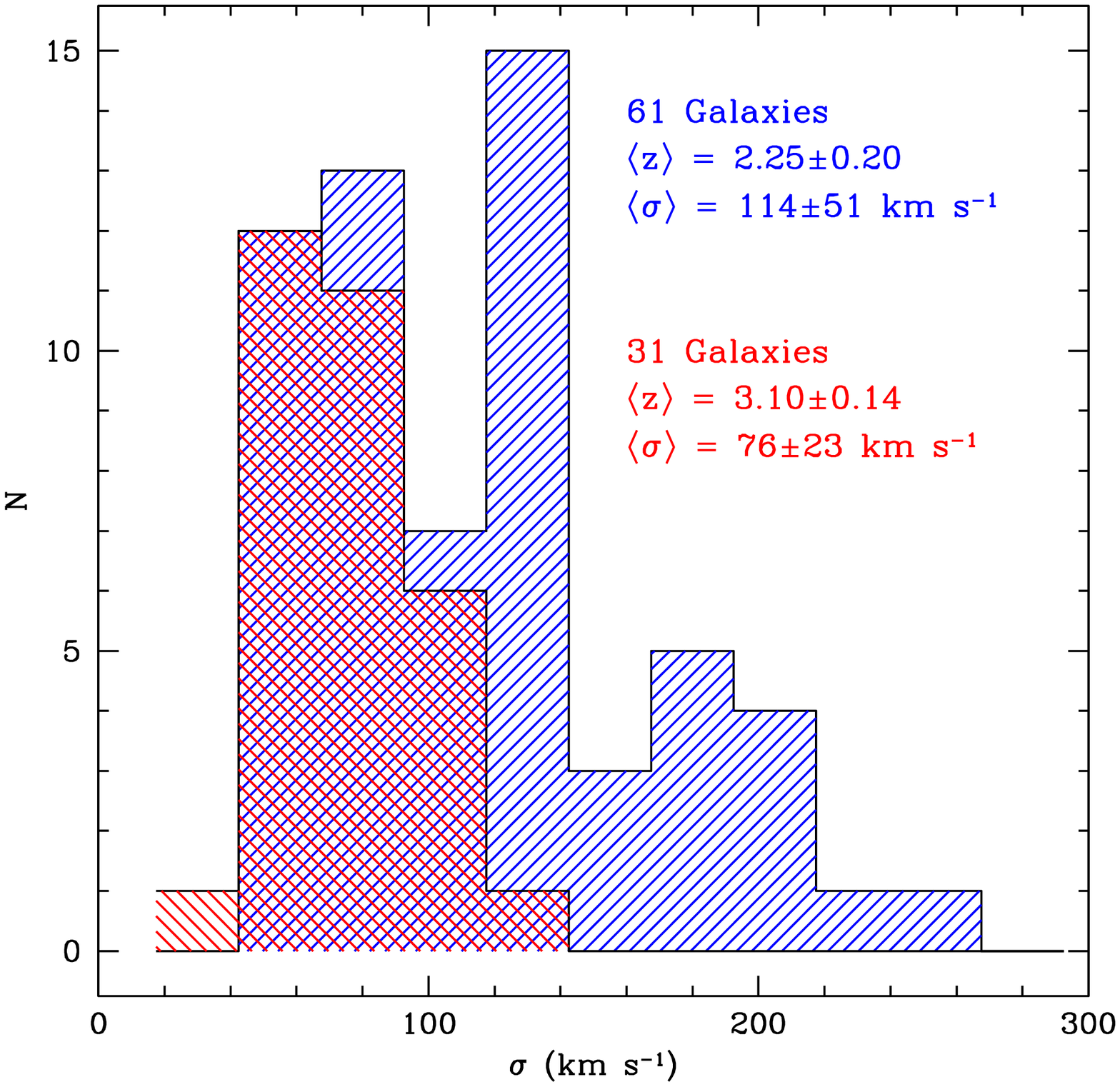}\includegraphics[width=0.5\textwidth]{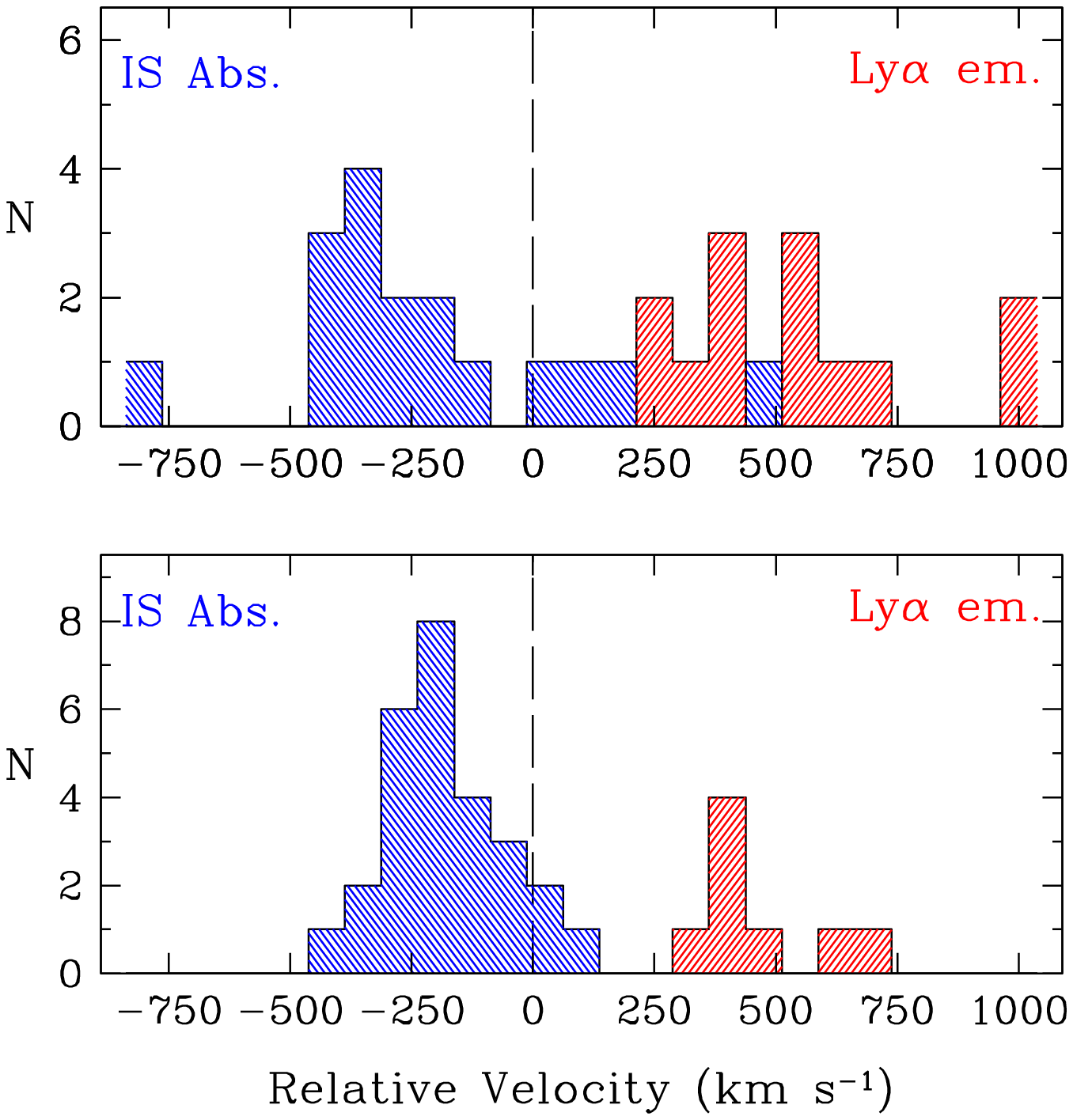}
\end{center}
\caption[]{{\bf Left:} A comparison  of the distribution of one-dimensional velocity
dispersions (as measured using H$\alpha$ or [OIII] lines) for a sample of LBGs
at $z\sim 3$ \cite{p01} (red) and $z\sim 2$ (blue). More than 50\% of the
$z \sim 2$ galaxies have $\sigma$ larger than the largest values observed
at $z \sim 3$. {\bf Right:} a comparison of the galactic outflow kinematics
of $z \sim 2$ (bottom) and $z \sim 3$ (top, \cite{p01}) galaxies.}
\label{kband}
\end{figure}
 
Here we focus on the kinematics and chemical abundances. Our first observations (\cite{erb03})
showed that the H$\alpha$ measurements could be made with a high success rate for the UV-selected
$z \sim 2$ galaxies, and that in a reasonably high fraction of the cases observed, spatially resolved
velocity shear was observed, such as might be expected in the case of rotating disks. The observability
of such velocity shear in high redshift galaxies is very seeing dependent \cite{erb04} but the 
overall line widths ($\sigma$) are also consistently larger than for the $z\sim 3$ galaxies
in the sample of \cite{p01} despite the fact that they were selected on very similar UV properties.
Comparisons of the $z\sim2$ and $z\sim 3$ samples are shown in Figure 4-- note that at $z \sim 2$, the
line widths are much larger, while the galactic outflow kinematics, characterized by the velocities
of the interstellar absorption lines and Lyman $\alpha$ emission with respect to the H$\alpha$-defined
redshift, are quite similar to the $z\sim3$ counterparts. The most straightforward interpretation
of both the 1-d line widths and the $K_s$ band properties (\S 3.1) is that the stellar masses of UV-selected
galaxies have increased considerably between the epochs corresponding to $z\sim3$ and $z \sim 2$.

\begin{figure}[tb]
\begin{center}
\includegraphics[width=0.5\textwidth]{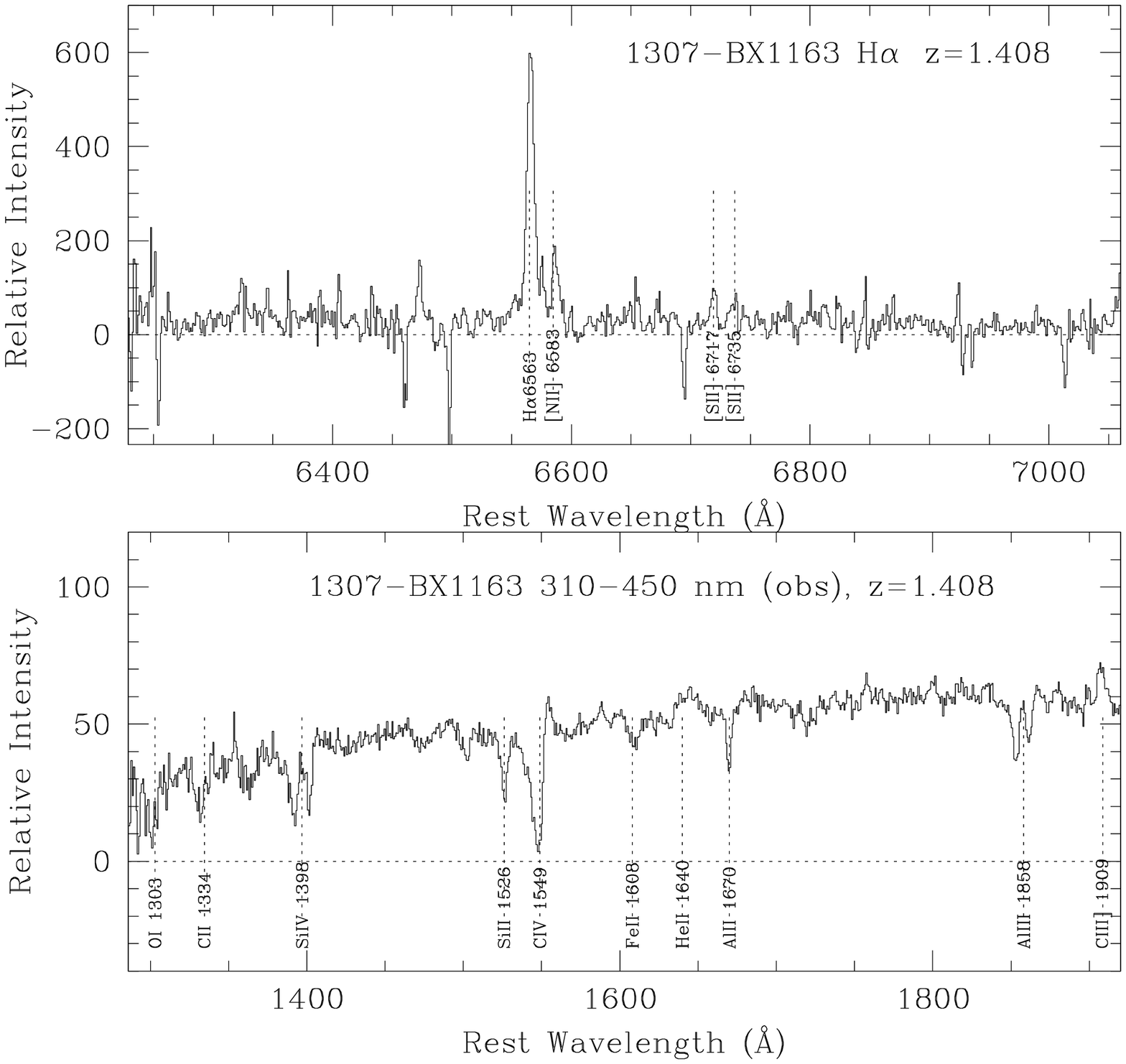}\includegraphics[width=0.5\textwidth]{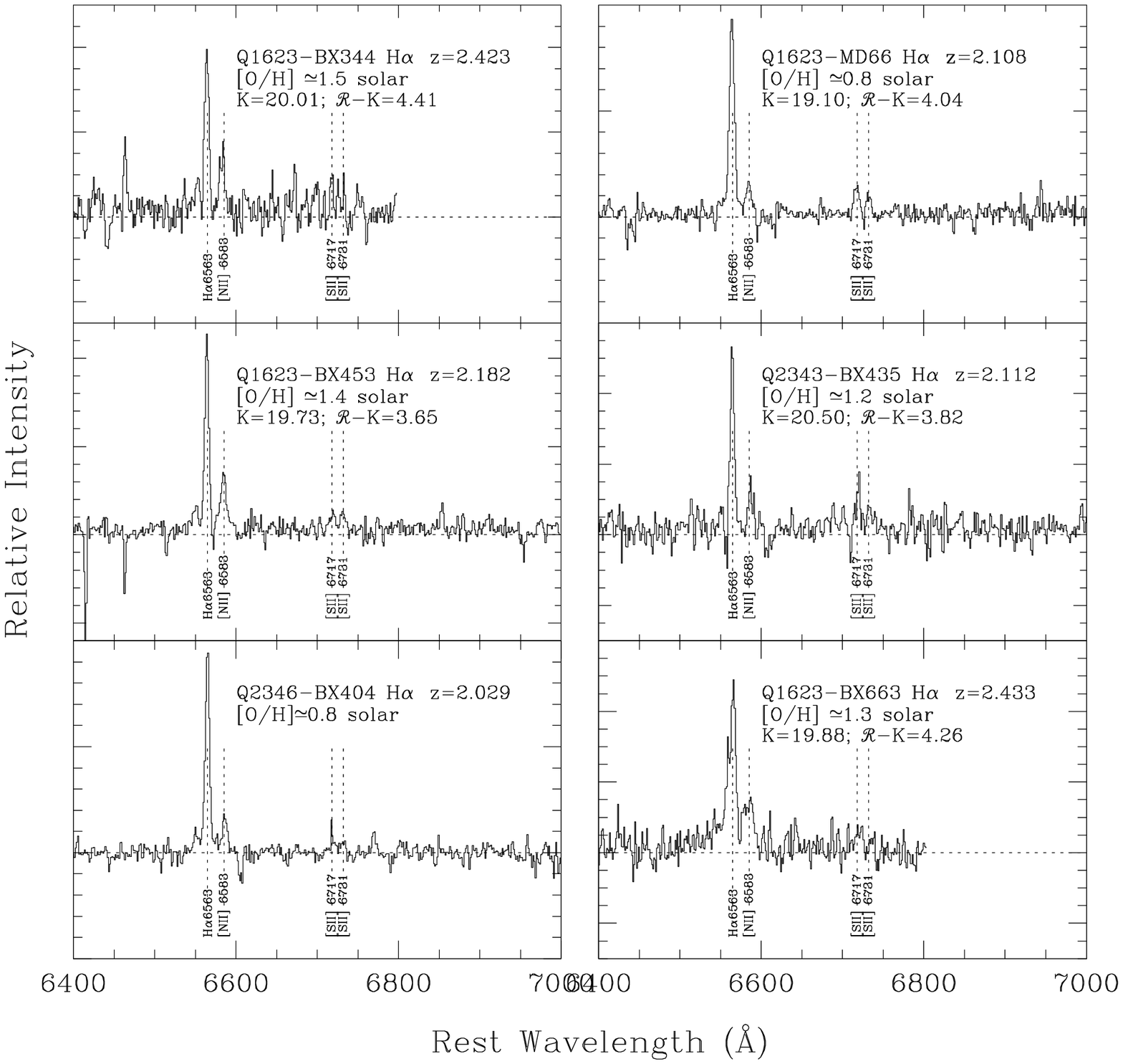}
\end{center}
\caption[]{{\bf Left:} LRIS-B (lower) and NIRSPEC (upper) spectra of a galaxy at $z=1.41$ illustrating 
the type of data from which detailed astrophysical properties of ``desert'' galaxies
can be extracted. {\bf Right:} Example NIRSPEC spectra of near-IR bright galaxies drawn
from our UV-selected sample at $z \sim 2$. Using the calibration of \cite{den02} these
have oxygen abundances that are consistent with being solar or greater.}
\label{abund}
\end{figure}

We have begun a program to obtain data suitable for measuring chemical abundances in 
the $z \sim 2$ galaxies, using both nebular indicators such as the [NII]/H$\alpha$ ratio
(\cite{den02}) and also the rest far-UV features recorded in the optical LRIS-B spectra (see,
e.g., Fig. 5a). Such observations allow significant insight into the physical nature of
the galaxies: for example, from the spectra in Fig. 5a, we can infer \cite{s04} that both
the nebular lines and the far-UV measurements indicate $\simeq$solar abundances, that the
high mass end of the IMF is consistent with Salpeter, that the velocity dispersion of the
HII regions of the galaxy is $\sigma = 126$ \kms, and that interstellar absorption lines
due to outflowing gas are blue-shifted by a bulk velocity of $\simeq 300$ \kms and
have a velocity width of $650$ \kms. In figure 5b, we show H$\alpha$/[NII] spectra of objects
selected to be either bright in the $K_s$ band or red in ${\cal R}-K_s$; for these objects,
which comprise a significant fraction of the UV-selected sample, there are indications
that the galaxies are typically solar metallicity or greater, even at $z \sim 2$; the
full results of such analysis will be presented in \cite{shap04}.    

\section{``Desert'' Results in the GOODS-N Field}  

While most of our effort to date on ``redshift desert'' galaxies has been
concentrated in specially chosen fields with multiple bright background
QSOs for the IGM component of the survey (not discussed here), we have also
worked in fields with extensive existing or planned multi-wavelength 
observations. One of these fields is GOODS-N (formerly known as HDF-N). 
This is one of two fields in which we have targeted significant numbers
of galaxies in both the $z\simeq 1.5-2$ and $z \simeq 2-2.5$ redshift range,
as well as at $z \sim 3$ (see Figure 6a). The existence of large spectroscopic
samples of $z > 1.4$ galaxies together with existing SCUBA, Chandra, and VLA  
images, as well as the very high quality GOODS/ACS images and planned
SIRTF/IRAC and  SIRTF/MIPS observations, will allow for a large amount of progress
and a deeper understanding of the energetics, masses, and stellar populations
of star forming galaxies at high redshifts. Here we make only a few comments.

There has been quite a bit of discussion about the overlap, or lack thereof, between
SCUBA galaxies detected at 850 $\mu$m and the UV-selected samples of LBGs
at $z \simgt 3$ (and now the large samples in the ``redshift desert''). In fact, it
turns out that there is indeed substantial overlap between the BX/BM samples
and the SCUBA/radio objects of \cite{chap03}, and the HDF region is no
exception. There are at least 4 objects in common between our BX/BM 
sample in GOODS-N (fig. 6a) and the radio-detected ``SMGs'' in the Chapman et al sample: 
3 of these have LRIS-B redshifts from our own survey, at $z=2.098$, $z=1.989$, and $z=1.865$. A glance at
fig. 6a shows that each of these is a member of one of the redshift desert ``spikes''
found from the LRIS-B spectroscopic results for BX/BM galaxies, qualitatively supporting
the claim (see Blain's contributions to these proceedings) that the SMGs are strongly
clustered objects. It is possible that the reason for the lack of overlap
between SCUBA sources and the $z \sim 3$ LBGs may be due to the fact that
the redshift distribution for SMGs peaks in the $z=2-2.5$ range, very much like optically
selected QSOs; $z > 3$ SMGs are relatively rare. The increase in overlap with SCUBA
sources is yet another piece of evidence for significant evolution in the sample of
UV-selected galaxies between $z \sim 3$ and $z \sim 2$.

\begin{figure}[tb]
\begin{center}
\includegraphics[width=0.45\textwidth]{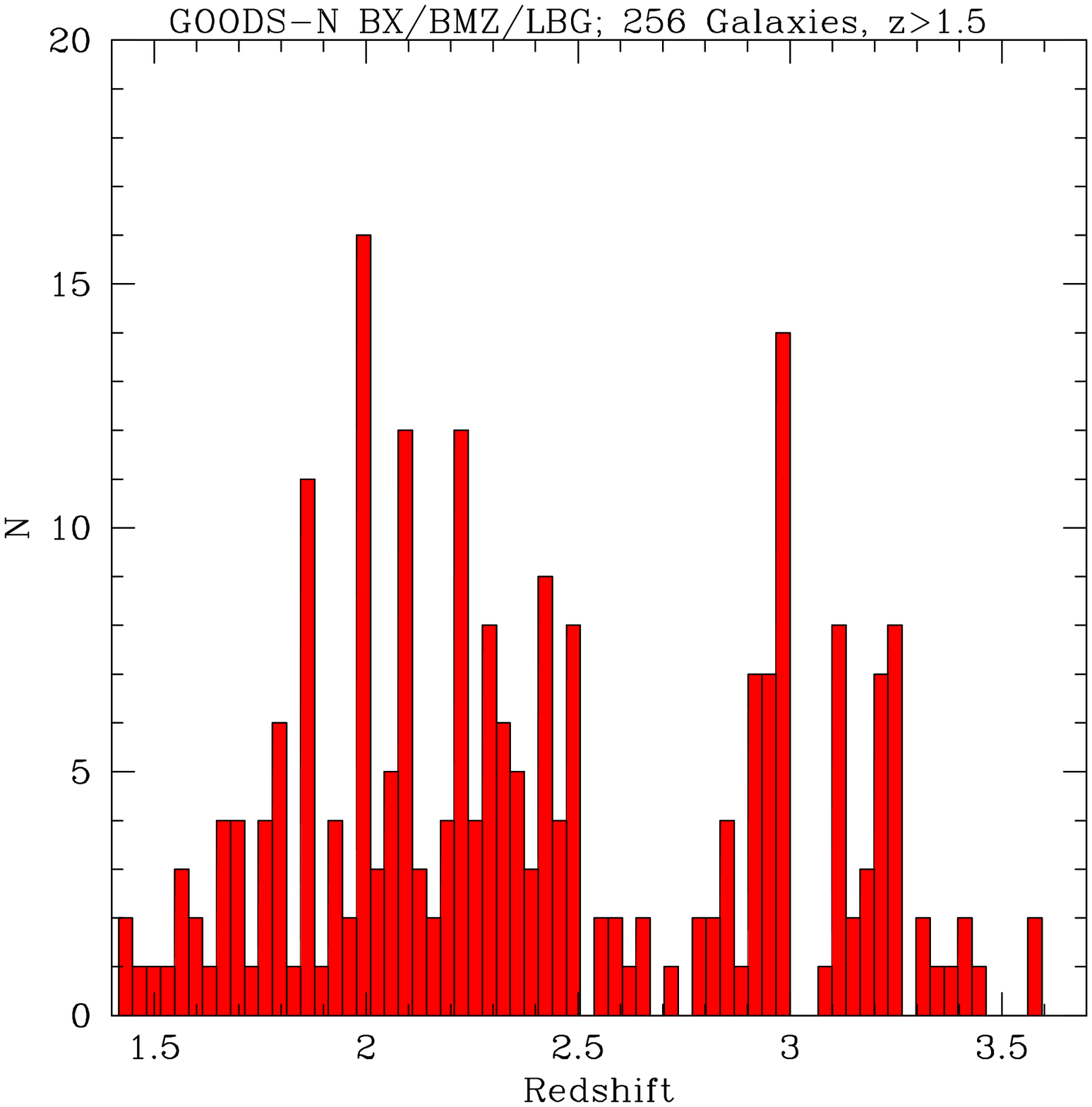}\includegraphics[width=0.45\textwidth]{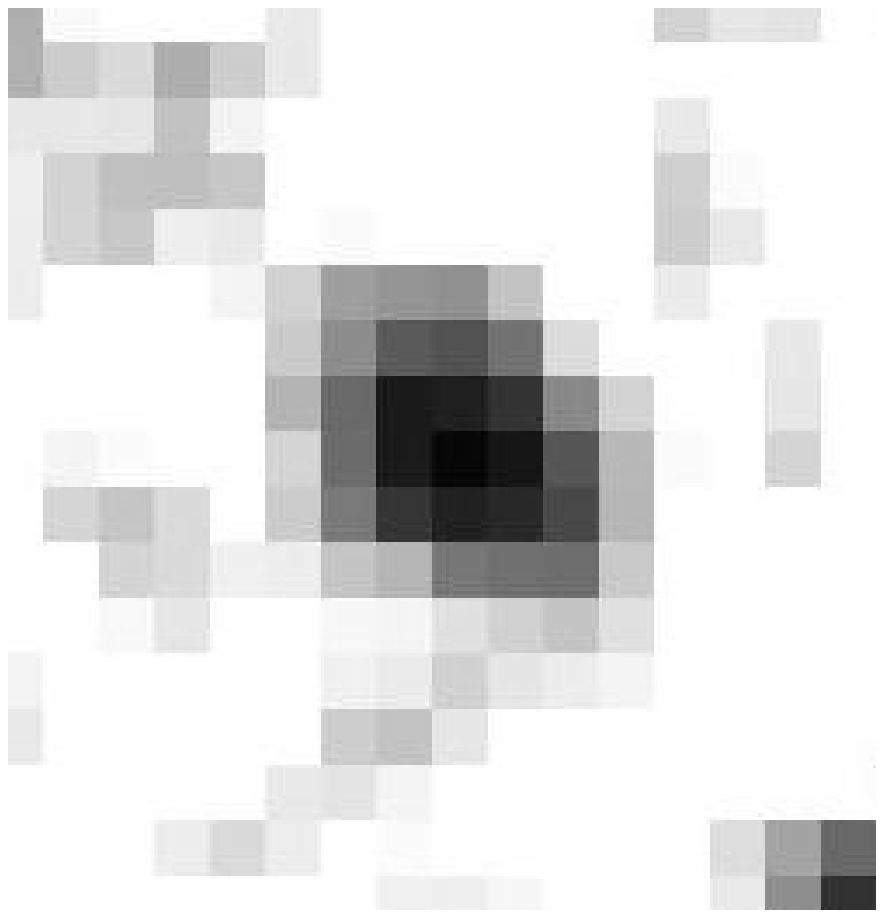}
\end{center}
\caption[]{{\bf Left:} Redshift histogram of 256 $z > 1.5$ GOODS-N galaxies selected using various
UV color selection techniques (those at $z < 2.6$ are based upon the BX and BM criteria). 
{\bf Right:} X-ray stack of 160 $z\sim 2$ UV-selected galaxies with spectroscopic redshifts.
The average X-ray flux implies an average star formation rate of $\simeq 50 \msun$ yr$^{-1}$, in
excellent agreement with both the average SFR estimated from the radio stack and that estimated
from the UV continuum slope. }
\label{goods}
\end{figure}

Very recently we have examined the average X-ray and radio emission from the spectroscopic $z \sim 2$ 
galaxy sample in the GOODS-N field via a stacking analysis \cite{rs04}, providing
a 3-way check on the average inferred star formation rate (and UV attenuation factor).
The stack of the X-ray image at the positions of 160 $z \sim 2$ galaxies
(after excluding all directly detected X-ray sources, among which are all 3 of the identified
sub-mm/UV galaxies) in the Chandra 2 Ms image \cite{alex03} is shown in fig. 6b.  
The stacked BX/BM galaxies are detected at the $\sim 10\sigma$ level.  A similar exercise has
been completed using the radio image of \cite{rich99}, and the estimated average star
formation rate has been computed from the ground-based optical photometry assuming the \cite{meu00}
relation between far-UV color and far-IR luminosity. The result is that all 3 means of estimating
SFR for the same sample suggest an {\it average} SFR of $\simeq 50 \msun$ yr$^{-1}$ ($h=0.7$,
$\Omega_m=0.3$, $\Omega_{\Lambda}=0.7$). The good agreement between UV, X-ray, and radio estimates
of star formation rates suggests that, on average, the locally calibrated relationships between
measurements at these wavelengths and the bolometric luminosity (i.e., SFR) still apply for starburst
type systems at high redshift. The results also imply that the average $z \sim 2$
UV selected galaxy in our spectroscopic sample has a far-IR luminosity of $\langle L_{\rm bol} \rangle
\simeq 2 \times 10^{11}$ L$_{\odot}$, which would qualify as a LIRG by the usual definition. 
The average implied UV extinction factor
is $\simeq 5$, very much in line with the value that has been suggested for $z \sim 3$ galaxies
\cite{s99}\cite{nan02}. With such typical luminosities, the spectroscopic $z \sim 2$ galaxies should be
easily detectable by SIRTF/MIPS in the 24$\mu$m band on an individual basis (rather than by stacking
a large number of objects). 

\section{Summary}

We have highlighted the fact that, with suitable optical photometric selection and
follow-up spectroscopy with a highly efficient UV/blue optimized spectrograph, the
``redshift desert'' ceases to exist. In fact, in many ways star forming galaxies at $z \sim 2$
offer more opportunities for measuring key physical quantities using current generation
telescopes than at any redshift beyond
the local universe, because of the simultaneous access to the rest-frame far-UV and rest-frame
optical spectra, and the observability of the galaxies' effects on the diffuse intergalactic medium.
With the addition of high quality space-based data, particularly from SIRTF, we should learn
a great deal about galaxies in this important cosmic epoch in the very near future. 

We have shown that there has been significant evolution in the kinematics and the optical/IR
colors of (identically) UV-selected galaxies between the epochs corresponding to $z \sim 3$ and $z \sim 2$.
Both of these observations are consistent with a significant evolution (a factor of $\sim 2$)
in the typical stellar mass of vigorously star forming systems over this redshift range.  There are
significant numbers of galaxies {\it still forming stars} at $z \sim 2$ which have already enriched
themselves to solar metallicity or greater, and which have stellar masses of $\sim 10^{11}$ $\msun$. 

More quantitative results in the areas discussed above, and analyses of the clustering properties,
luminosity functions, and the relationship to and interaction with the diffuse IGM are all
in preparation. We would like to thank the David and Lucile Packard Foundation and the
US National Science Foundation for support for this work.

%

\end{document}